\documentclass[12pt]{article}
\usepackage[cp1251]{inputenc}
\usepackage[russian]{babel}
\usepackage{amssymb}

\newcommand{\bm}[1]{\mbox{\boldmath{$#1$}}}
\usepackage{cite}
\begin{document}
\thispagestyle{empty}

\begin{center}
{\bf ON THE STABILITY OF TRIANGULAR LAGRANGIAN POINTS IN THE SPATIAL RESTRICTED THREE-BODY PROBLEM}
\end{center}

\vspace{0.2cm}

\begin{center}
{\large         S. P. Sosnitskii}\\
{\small
{\it Institute of Mathematics of National Academy of Sciences of Ukraine,\\
Tereshchenkivs'ka  str 3, 01601,MSP, Kyiv--4, Ukraine,\\
E-mail address: sosn@imath.kiev.ua}} \\
\end{center}

\vspace{0.2cm}

\noindent {\bf Abstract.} {\small{In the present paper, which is a development of an earlier study by the author \cite{Sosnitskii08}, we consider the stability of triangular libration points in the spatial circular restricted three-body problem and  improve the result of author's work \cite{Sosnitskii08}. Unlike \cite{Sosnitskii08}, where the instability of libration points was established on the base of reduced approximate equations, we succeeded in this paper to use a new approach  that made it possible to prove the instability of triangular libration points on the base of a closed complete system of equations. The relationship between the Lyapunov stability and Birkhoff stability (formal stability) is also discussed}.


\section{Introduction}

Although the problem of stability of triangular libration points in the circular restricted three-body problem  was already posed in the 19th century \cite{Gascheau,Routh}, this problem of stability is not completely solved until now.
Seemingly, it may be supposed that the KAM theory created by Kolmogorov \cite{Kolmogorov}, Arnold \cite{Arnold1}, and Moser \cite{Moser} could solve the problem, but it solves the problem only in the case of the planar circular restricted problem \cite{Leont,Deprit,Mark,Sokol,Arnold2}.
As for the stability in the spatial circular problem, the KAM theory applied to it does not lead to the desired result.

Below, continuing our research on the stability of triangular libration points in the restricted circular problem \cite{Sosnitskii08,Sosnitskii09}, we find a new source in the depths of the Jacobi integral.
In particular, using the Jacobi integral, we arrive at a special form of equations of the perturbed motion in a neighborhood of the libration points. These equations make it possible to solve the problem of stability effectively in the frame of the existing methods of the theory of stability.

\section{Equations of Motion for the Circular Restricted Three-Body Problem}

Let us consider the spatial circular restricted three-body problem in the case where the vectors ${\bf r}_{1}$ and ${\bf r}_{2}$, being solutions to the two-body problem, correspond to the circular orbits of points with masses $m_1$ and $m_2$.
Passing over to the relative lengths of vectors \cite{Sosnitskii08}
$$
{\bm\rho}_i=\frac{{\bf r}_{i}}{|{\bf r}_{12}|}, \eqno(1)
$$
where $|{\bf r}_{12}|=|{\bf r}_{12}|_0=\mbox{const}$, we can write the equations of motion as follows:
$$
{{\bm \rho}_{1}}''= {\mu} \frac{{\bm \rho}_{12}} {|{\bm \rho}_{12}|^3},
$$
$$
{{\bm \rho}_{2}}''=-{(1-\mu)}\frac{{\bm \rho}_{12}} {|{\bm \rho}_{12}|^3},
$$
$$
{{\bm \rho}_{3}}''= -{(1-\mu)} \frac{{\bm \rho}_{13}} {|{\bm \rho}_{13}|^3} - \mu \frac {{\bm \rho}_{23}}{|{\bm \rho}_{23}|^3}. \eqno(2)
$$
Here, ${\bm \rho}_{ij} = {\bm\rho}_{j} - {\bm\rho}_{i}$ $(i,j = 1,2,3)$, and the prime sign denotes the differentiation over dimensionless time
$$
\tau= \frac{\sqrt{G(m_1+m_2)}}{{|{\bf r}_{12}|_0}^{3/2}}t,
$$
where  $G>0$ is the gravitational constant and
$$
\mu=\frac{m_2}{m_1+m_2}, \ \ 0<\mu\leq \frac{1}{2}.
$$

System (2) can be rewritten in the form
$$
{{\bm \rho}_{12}}''= -\frac{{\bm \rho}_{12}} {|{\bm
\rho}_{12}|^3},
$$
$$
{{\bm \rho}_{3}}''= {-(1-\mu)} \frac{{\bm \rho}_{13}} {|{\bm
\rho}_{13}|^3} - \mu \frac {{\bm \rho}_{23}}{|{\bm \rho}_{23}|^3}.   \eqno(3)
$$

If we use a coordinate system rotating with unit angular velocity around an axis perpendicular to the plane of rotation of two massive bodies, then the second vector equation of system (3) in this coordinate system takes the form \cite{Szebehely67}:
$$
x''-2y'=x-(1-\mu)\frac{x-\mu}{\rho_{13}^3}-\mu\frac{x+1-\mu}{\rho_{23}^3},
$$
$$
y''+2x'=y-(1-\mu)\frac{y}{\rho_{13}^3}- \mu\frac{y}{\rho_{23}^3},
$$
$$
z''=-(1-\mu)\frac{z}{\rho_{13}^3}- \mu\frac{z}{\rho_{23}^3}. \eqno(4)
$$
Here, $\rho_{13}=|{\bm\rho}_{13}|, \quad \rho_{23}=|{\bm\rho}_{23}|$,
$$
{\bm\rho}_{13}^2
=(x-\mu)^2+y^2+z^2,\quad{\bm\rho}_{23}^2
=(x+1-\mu)^2+y^2+z^2,
\eqno(5)
$$
where $(x,y,z)$ are coordinates of the small particle relative to the rotating coordinate system.

Next, we denote $(x,y,z)^T={\bf r}$, $(\tilde{x},\tilde{y},z)^T={\bm\rho}_{3}$, where $(\tilde{x},\tilde{y},z)$ are  coordinates of the small particle with respect to the inertial coordinate system. Then we arrive at the equality ${\bf r}^2={\bm\rho}_{3}^2$.
Next, it is convenient to rewrite equalities (5) in the form
$$
{\bm\rho}_{13}^2
=-2\mu x+\mu^{2}+{\bf r}^2,\quad{\bm\rho}_{23}^2
=2(1-\mu)x +(1-\mu)^2+{\bf r}^2.
\eqno(6)
$$
This implies
$$
{\bf r}^2={\bm\rho}_{3}^2 =-\mu(1-\mu) +(1-\mu){\bm\rho}_{13}^2+\mu{\bm \rho}_{23}^2.
\eqno(7)
$$
In connection with (7), we also note that the following equality is valid:
$$
{\bm\rho}_{3}'^2
=-\mu(1-\mu) +(1-\mu){\bm\rho}_{13}'^2+\mu{\bm \rho}_{23}'^2.
\eqno(8)
$$

The Jacobi integral for system (4) has the form
$$
{x'}^2+{y'}^2+{z'}^2-\left(x^{2}+y^{2}\right)-\frac{2(1-\mu)}{|{\bm \rho}_{13}|}- \frac{2\mu}{|{\bm\rho}_{23}|}=2h, \quad h=\mbox{const}. \eqno(9)
$$

Along with the equations of motion in the form (4), for our further purposes, we will also use the following distance equations obtained in \cite{Sosnitskii08}:
$$
{\rho_{13}^2}''=2E_{13}+\frac{2}{\rho_{13}}+\mu
\left[\frac{2}{\rho_{13}}+
(\rho_{23}^2-\rho_{13}^2-1)+\frac{1}{\rho_{23}} \left(\frac
{1-\rho_{13}^2}{\rho_{23}^2}-1\right)\right],
$$
$$
{\rho_{23}^2}''=2E_{23}+\frac{2}{\rho_{23}}+(1-\mu)
\left[\frac{2}{\rho_{23}}+(\rho_{13}^2-\rho_{23}^2-1)+
\frac{1}{\rho_{13}}\left(\frac
{1-\rho_{23}^2}{\rho_{13}^2}-1\right)\right],
$$
$$
E_{13}'={-\mu}\left[(\rho_{13}^2)'\left(1-\frac{1}{\rho_{13}^3}\right)-
(\rho_{23}^2)'\left(1-\frac{1}{\rho_{23}^3}\right)+2y
\left(1-\frac{1}{\rho_{23}^3} \right)\right],
$$
$$
E_{23}'=(1-\mu)\left[(\rho_{13}^2)'\left(1-\frac{1}{\rho_{13}^3}\right)-
(\rho_{23}^2)'\left(1-\frac{1}{\rho_{23}^3}\right)+2y
\left(1-\frac{1}{\rho_{13}^3} \right)\right],
$$
$$
2y'=E_{23}- E_{13}+ \rho_{13}^2- \rho_{23}^2+ \frac{2}{\rho_{23}}-
\frac{2}{\rho_{13}}, \eqno(10)
$$
where
$$
E_{13}=v_{13}^2-\frac{2}{\rho_{13}},\quad
E_{23}=v_{23}^2-\frac{2}{\rho_{23}},\quad v_{13}={\bm \rho}_{13}', \quad v_{23}={\bm \rho}_{23}'.   \eqno(11)
$$

Based on equations (10), the energy integral can be represented as follows\cite{Sosnitskii08}:
$$
-\mu(1-\mu)+(1-\mu)E_{13}+\mu E_{23}-2({x}{y}'-{y}{x}'+
x^{2}+y^{2})=2h. \eqno(12)
$$

It is known that systems (4) and (10) have equilibrium solutions, which are commonly called libration points or Lagrange points. Restricting further consideration to only triangular libration points $L_4$ and $L_5$, we turn to the study of their stability.

\section {Equations of Perturbed Motion in a Neighborhood of Libration Points}

When studying the stability of libration points (stationary Lagrange triangles), we will mainly use equations (10).

In accordance with the choice of dependent variables, the stationary Lagrange triangle in system (10) corresponds to the solution
$$
\rho_{13}^2=\rho_{23}^2=\rho_0^2=1, \quad E_{13}=E_{23}=E^0=-1.
\eqno(13)
$$
For system (4), we have
$$
x^0=-\frac{1}{2}+\mu, \quad (y^0)^2=\frac{3}{4},\quad z^0=0.
\eqno(14)
$$
(see \cite{Sosnitskii08} for details).

For perturbations of the stationary solution of system (10), we introduce the notation
$$
x_1=\rho_{13}^2-1, \quad x_2=\rho_{23}^2-1,\quad y_1=E_{13}+1,\quad y_2=E_{23}+1,\quad  \eta=y-y^0.     \eqno(15)
$$
Similarly, for system (4), we have
$$
\xi=x-x^0,\quad \eta=y-y^0,\quad \zeta=z-z^0. \eqno(16)
$$

Just as there is a correspondence between squared distances $\rho_{13}^2$ and $\rho_{23}^2$ and coordinates $x, y, z$ of the small particle in the moving reference frame, we also have the following correspondence between their small displacements in the vicinity of the libration points $L_4, L_5$ \cite{Sosnitskii08}:
$$
x_1=-\xi + 2y^0\eta+{\bf u}^2,\quad x_2=\xi + 2y^0\eta+{\bf u}^2, \quad
{\bf u}=(\xi,\eta,\zeta)^T,
\eqno(17)
$$
$$
\xi=\frac{1}{2}(-x_1+x_2), \quad  {\bf x}=(x_1,x_2)^T,
$$
$$
\eta=\frac{1}{3}y^0\left\{(x_1+x_2)-\frac{2}{3}(x_1^2+x_2^2-x_1x_2)-2\zeta^2+
O\left[({\bf x}^2+\zeta^2)^{3/2}\right]\right\}.
\eqno(18)
$$

In a small neighborhood of the libration points $L_4$ and $L_5$, we arrive at the equations of perturbed motion \cite{Sosnitskii08}
$$
{x_1}''=2y_1-(1+3\mu)x_1+\frac{3}{2}\mu x_2+  O({\bf x}^2),
$$
$$
{x_2}''=2y_2+\frac{3}{2}(1-\mu)x_1-(4-3\mu)x_2+ O({\bf x}^2),
$$
$$
{y_1}'=-3\mu y^0x_2 + O\left({\bf x}^2+{\bf x}'^2+\eta^2\right),
$$
$$
{y_2}'=3(1-\mu)y^0x_1+O\left({\bf x}^2+{\bf x}'^2 + \eta^2\right),
$$
$$
{\eta}'=\frac{1}{2}(y_2-y_1)+ x_1-x_2+ O({\bf x}^2), \eqno(19)
$$
which correspond to system (10). Similarly, for the equations of perturbed motion in the form (4) we have
$$
\xi''-2\eta'=\frac{3}{4}\xi-\frac{3}{2}y^0(1-2\mu)\eta
+O(||{\bf u}||^2),
$$
$$
\eta''+2\xi'=-\frac{3}{2}y^0(1-2\mu)\xi+\frac{9}{4}\eta+O(||{\bf u}||^2),
$$
$$
\zeta''=-\zeta+O(||{\bf u}||^2). \eqno(20)
$$

Further, we use the previously obtained equalities \cite{Sosnitskii08}
$$
E_{13}=2T+2(xy'-yx')-2\mu y'+ x^2+y^2  - 2\mu x + \mu^2-\frac{2}{\rho_{13}},
\eqno(21)
$$
$$
E_{23}=2T+2(xy'-yx')+2(1-\mu) y'+x^2+y^2  + 2(1-\mu)x +(1- \mu)^2-\frac{2}{\rho_{23}},
\eqno(22)
$$
which, taking into account (13)--(18), can be conveniently rewritten in the form
$$
y_1
=2x_1+\frac{2}{3}y_0(x_1'-2x_2')+\zeta'^2-\zeta^2+\frac{2}{3}y^0(\zeta^2)'
+f_{1}\left({\bf x},{\bf x}'\right)+\Delta_{1}({\bf x},{\bf x}',\zeta^{2},{\zeta^{2}}'),
\eqno(23)
$$
$$
y_2=2x_2+\frac{2}{3}y_0(2x_1'-x_2')+\zeta'^2-\zeta^2-\frac{2}{3}y^0(\zeta^2)'+f_{2}\left({\bf x},{\bf x}'\right)+\Delta_{2}({\bf x},{\bf x}',\zeta^{2},{\zeta^{2}}'), \eqno(24)
$$
where  $f_{i}=O(||{\bf q^{*}}||^2)$, ${\bf q^{*}}=({\bf x},{\bf x}')^T$, $\Delta_{i}=O(||{\bf q}||^3)=O({\bf x}^2+{\bf x}'^2+\zeta^2+|{\zeta^2}'|)^{3/2}$, $(i=1,2)$.
Then the system of equations (19) can also be represented in the form
$$
{x_1}''=\frac{4}{3}y_0(x_1'-2x_2')+ 3(1-\mu)x_1+\frac{3}{2}\mu x_2+2(\zeta'^2-\zeta^2)+\frac{4}{3}y^0(\zeta^2)'+
O(||{\bf q^{*}}||^2)+O(||{\bf q}||^3),
$$
$$
{x_2}''=\frac{4}{3}y_0(2x_1'-x_2')+\frac{3}{2}(1-\mu)x_1+3\mu x_2+2(\zeta'^2-\zeta^2)-\frac{4}{3}y^0(\zeta^2)'+
O(||{\bf q^{*}}||^2)+O(||{\bf q}||^3),
$$
$$
\zeta''=-\zeta+O({\bf x}^2+\zeta^2),\eqno(25)
$$
thereby eliminating the redundant coordinate.

For a perturbed motion, we consider the Jacobi integral in one of the following forms:
$$
(1-\mu)y_{1}+\mu y_{2}+(1-2\mu)\eta'+2y^0\xi'+2(1-2\mu)\xi-4y^0\eta-2(\xi \eta'-\eta \xi'+
\xi^{2}+\eta^{2})=2h^*, \eqno(26)
$$
$$
(1-\mu)y_{1}+\mu y_{2}-\frac{2}{3}y^{0}[(1+\mu)x_1'- (2-\mu)x_2']-2[(1-\mu)x_1+\mu x_2]-
$$
$$
-\frac{2}{3}y^{0}(1-2\mu){\zeta^2}'+2\zeta^2+O(||{\bf q^{*}}||^2)+O(||{\bf q}||^3) =2h^*,\eqno(27)
$$
$$
\frac{1}{3}(x_{1}'^{2}+x_{2}'^{2}-x_{1}'x_{2}')+{\zeta'}^2-
\frac{3}{4}[(1-\mu)x_1^2+\mu x_2^2]+\zeta^2+O(||{\bf q}||^3)=2h^*.\eqno(28)
$$
As we can see, its most compact form is (26). However, for our further research, all three forms of notation are important.

\section{On the Instability of Libration Points}

In what follows, when studying the stability of Lagrangian triangles, we will transform the system of equations (19) to a form that would allow us to perform the stability analysis as simple as possible.

\medskip

{\bf Lemma 1}. {\it System (19) can be represented as
$$
{x_1}''=2v-(1-\mu)x_1 - \frac{5}{2}\mu x_2-4\mu\eta'+O\left(||{\bf x}||^2\right),
$$
$$
{x_2}''=2v -\frac{5}{2}(1-\mu)x_1 - \mu x_2+4(1-\mu)\eta'+ O\left(||{\bf x}||^2\right).
$$
$$
{v}'=3\mu (1-\mu)y^0(x_1-x_2)+\gamma\left({\bf x},{\bf x}',\zeta^2\right),
$$
$$
\zeta''=-\zeta+O({\bf u}^2),\ {\bf u}=(\xi,\eta,\zeta)^T,\eqno(29)
$$
where $\gamma=O({\bf x}^2+{\bf x}'^2+\zeta^2)$.
}
\medskip

{\bf Proof}. Representing equality (12) in the form
$$
-\mu(1-\mu)+(1-\mu)E_{13}+\mu E_{23}-2({x}{y}'-{y}{x}'+
x^{2}+y^{2})=2T-2U
$$
and resolving it with respect to $2T$, we substitute the obtained value into (21) and (22).
As a result, we have
$$
E_{13}=(1-\mu)E_{13}+\mu
E_{23}+2\mu\left(\frac{1}{\rho_{23}}-\frac{1}{\rho_{13}}\right)
-2\mu y' - \mu(\rho_{23}^2-\rho_{13}^2),\eqno(30)
$$
$$
E_{23}=(1-\mu)E_{13}+\mu
E_{23}+2(1-\mu)\left(\frac{1}{\rho_{13}}-\frac{1}{\rho_{23}}\right)
$$
$$
+ 2(1-\mu) y'+(1-\mu)(\rho_{23}^2-\rho_{13}^2).\eqno(31)
$$
Taking into account equalities (13)--(18), based on (30) and (31) we obtain
$$
y_{1}=v-2\mu(x_{2}-x_{1})-2\mu\eta'+ O\left(||{\bf x}||^2\right),\eqno(32)
$$
$$
y_{2}=v+2(1-\mu)(x_{2}-x_{1})+2(1-\mu)\eta'+O\left(||{\bf x}||^2\right),\eqno(33)
$$
where
$$
v=(1-\mu)y_{1}+\mu y_{2}. \eqno(34)
$$

After this, if we replace $y_{1}$ and $y_{2}$ in the first two equations of system (19) by their values determined by equalities (32)--(34), then we obtain the first two equations of system (29).
Let us complete the resulting equations with the last equation of system (25) and the equation formed using the third and fourth equations of system (19). Then we arrive at system (29). $\Box$

\medskip

{\bf Lemma 2.} {\it In the case of circular restricted three-body problem, the following equality is satisfied:
$$
{\zeta}''=4(h-\zeta^{2})+O(||{\bf q^{*}}||^2)+O(||{\bf q}||^3).\eqno(35)
$$
}
\medskip

{\bf Proof.} By last equality of the system (29), we obtain
$$
{\zeta^2}''=2(\zeta'^2-\zeta^2)+O(\zeta{\bf u}^2),\ {\bf u}=(\xi,\eta,\zeta)^T.\eqno(36)
$$
Taking into account (36), we rewrite the Jacobi integral in the form (28) as follows:
$$
\frac{1}{3}(x_{1}'^{2}+x_{2}'^{2}-x_{1}'x_{2}')-
\frac{3}{4}[(1-\mu)x_1^2+\mu x_2^2]+\frac{1}{2}{\zeta^2}''+O({\bf x}^2+{\bf x}'^2+\zeta^2+|{\zeta^2}'|)^{3/2}=2(h^*-\zeta^2).\eqno(37)
$$
Based on the equality (37), we conclude that the lemma 2 is true. $\Box$

\medskip

{\bf Theorem}. {\it In the framework of the spatial circular restricted three-body problem, the libration points $L_{4}$ and $L_{5}$ are unstable in the sense of Lyapunov}.

\medskip

{\bf Proof}. In what follows, without loss of generality, we assume that $27\mu(1-\mu)\leq 1$.

Let us rewrite the Jacobi integral in the form (27) as follows:
$$
2v-\frac{4}{3}y^{0}[(1+\mu)x_1'- (2-\mu)x_2']-4[(1-\mu)x_1+\mu x_2]-
$$
$$
-\frac{4}{3}y^{0}(1-2\mu){\zeta^2}'+2(\zeta^2-h^*)+O(||{\bf q^{*}}||^2)+O(||{\bf q}||^3) =2(h^*-\zeta^2).\eqno(38)
$$
In  virtue of  lemma 2, we rewrite equality (38) in the form
$$
2v-\frac{4}{3}y^{0}[(1+\mu)x_1'- (2-\mu)x_2']-4[(1-\mu)x_1+\mu x_2]-
$$
$$
-\frac{4}{3}y^{0}(1-2\mu){\zeta^2}'+2(\zeta^2-h^*)+O(||{\bf q^{*}}||^2)+O(||{\bf q}||^3)=\frac{{\zeta^2}''}{2}.\eqno(39)
$$
Introducing the new variable
$$
\tilde{\zeta}=\zeta^2-h^*, \eqno(40)
$$
equality (39) is further conveniently represented in the form
$$
{\tilde{\zeta}}''=4v-\frac{8}{3}y^{0}[(1+\mu)x_1'- (2-\mu)x_2']-8[(1-\mu)x_1+\mu x_2]-
$$
$$
-\frac{8}{3}y^{0}(1-2\mu){\tilde{\zeta}}'+4\tilde{\zeta}+O(||{\bf q^{*}}||^2)+O(||{\bf q}||^3).\eqno(41)
$$

Taking into account the dependence of the right-hand sides of systems (25) and (29) on the variable ${\zeta}^2$, the introduction of the variable $\tilde{\zeta}$ seems to be natural.
Now  by replacing the last differential equation of the system (29) by a  differential equation  (41),  we transform system (29) into the following one:
$$
{x_1}''=2v-(1-\mu)x_1 - \frac{5}{2}\mu x_2-\frac{4}{3}\mu
y^0({x_1}+{x_2}-2\tilde{\zeta})'
+O(||{\bf q^{*}}||^2)+\gamma_{1}\left({\bf x},{\bf x}',\tilde{\zeta},{\tilde{\zeta}}',h^*\right),
$$
$$
{x_2}''=2v -\frac{5}{2}(1-\mu)x_1 - \mu x_2+\frac{4}{3}(1-\mu)
y^0({x_1}+{x_2}-2\tilde{\zeta})'+O(||{\bf q^{*}}||^2)+\gamma_{2}\left({\bf x},{\bf x}',\tilde{\zeta},{\tilde{\zeta}}',h^*\right),
$$
$$
{v}'=3\mu (1-\mu)y^0(x_1-x_2)+O(||{\bf q^{*}}||^2)+\gamma_{3}\left({\bf x},{\bf x}',\tilde{\zeta},h^*\right),
$$
$$
{\tilde{\zeta}}''=4v-\frac{8}{3}y^{0}[(1+\mu)x_1'- (2-\mu)x_2']-8[(1-\mu)x_1+\mu x_2]-
$$
$$
-\frac{8}{3}y^{0}(1-2\mu){\tilde{\zeta}}'+4\tilde{\zeta}+O(||{\bf q^{*}}||^2)+O(||{\bf \tilde{q}}||^3).\eqno(42)
$$
where $\gamma_{1},\gamma_{2}=O(||{\bf \tilde{q}}||^3)=O({\bf x}^2+{\bf x}'^2+|\tilde{\zeta}|+|{\tilde{\zeta}}'|+|h^*|)^{3/2}$, $\gamma_{3}=O({\bf x}^2+{\bf x}'^2+|\tilde{\zeta}|+|h^*|)^{3/2}$.

So, based on equations (29) and (41), we managed to construct a system of equations that is closed with respect to the variables $x_{1}$, $x_{2}$. $v$, and $\tilde{\zeta}$. Unlike the system of equations (25), the system (42) contains a redundant coordinate. However, as we will see below, this system  is more preferable for stability study than system (25). Indeed,
the related characteristic equation for the linear approximation of system (42) is representable  in the form
$$
\left[\frac{3}{4}\lambda^3+y^0(1-2\mu)\lambda^2-(5-8\mu+8\mu^2)\lambda+8(1-\mu+\mu^2)+4y^0(1-2\mu)\right]\times
$$
$$
\times\left[\lambda^4+\lambda^2+\frac{27}{4}\mu(1-\mu)\right]=0.\eqno(43)
$$
Under the conditions of theorem,  the equation
$$
\frac{3}{4}\lambda^3+y^0(1-2\mu)\lambda^2-(5-8\mu+8\mu^2)\lambda+8(1-\mu+\mu^2)+4y^0(1-2\mu)=0 \eqno(44)
$$
has all three  roots  with a non-zero real part. Consequently, the system (42) has a non-zero characteristic Lyapunov exponents.

Since the original system (3) is conservative, we arrive at the instability of the equilibrium state ${\bf x}={\bf x}'={\bf 0}, v=\tilde{\zeta}={\tilde{\zeta}}'=0$ of system (42) regardless of the sign of the  real parts  of the roots of equation (44), which is  determined by the sign of $y^0$.

The dimension of the phase space of the original system is six. Now, if we take into account the structure of the roots of equation (44), we come to the conclusion that the restriction of system (42) to six-dimensional space leaves its equilibrium position unstable.  As a sequence, this  causes the instability of the libration points $L_{4}$ and $L_{5}$.  $\Box$

The conservativity of the original system (3) and the structure  of the roots of the characteristic equation of system (42)
imply the existence of trajectories that are attracted to the libration points $L_{4}$ and $L_{5}$ for both $\tau\to {\infty}$ and $\tau\to -{\infty}$.

Thus, by a chain of transformations, we have come from system (19) to system (42), which has a non-zero characteristic Lyapunov exponents. It is obvious that the key point in the proof of the theorem is equation (41), which we have obtained using the Jacobi integral and lemma 2.  One can notice that the   last equation of system (29) contains the variable $\zeta$, which has the first order of smallness in the neighborhood of the considered libration points $L_{4}$ and $L_{5}$. On the other hand,  the equation (41) (last equation of system (42)) contains the variable $\tilde{\zeta}$. This variable, due to (40),  in the neighborhood of $L_{4}$ and $L_{5}$ in system (42) is a quantity of the second order of smallness. Since, as we already mentioned above, the right-hand sides of the equations under consideration contain ${\zeta}^2$, then by including the variable $\tilde{\zeta}$ in system (42),  apparently more fully and accurately, we reflected the internal structure of the spatial circular restricted three-body problem.

In contrast to \cite{Sosnitskii08}, where the stability of triangular libration points in the spatial circular restricted three-body problem  was considered for $2h^*\leq 0$, the approach proposed in this paper does not impose any restrictions on the constant defined for the Jacobi integral.

\section {Conclusion}

Summarizing the above, we can note that the study of the stability of stationary Lagrangian triangles have been reduced to a simpler system due to a successful choice of dependent variables and the use of an internal resource of the Jacobi integral.
The resulting system made it possible to prove the instability of Lagrangian triangles by using known methods. It is clear that the successful realization of this goal was facilitated by the intrinsic features of the structure of the spatial circular restricted problem, in particular, of the Jacobi integral in the form (27).

As follows from the above proved theorem and the method of its proof, the instability of the triangular libration points $L_4$ and $L_5$ in the spatial circular restricted three-body problem is in no way related to the problem of resonances.
Probably, the problem of resonances reflects rather the characteristic features of the KAM theory itself, and in this sense, the spatial circular restricted three-body problem  is undoubtedly interesting as an example of a system with three degrees of freedom, in which the formal stability of an equilibrium state (the Birkhoff stability \cite{Birk}) and its Lyapunov instability \cite{Lya} can coexist.

\newpage

\renewcommand \refname{References}

\end{document}